\documentclass{cpbtex}

\usepackage{amsmath}
\usepackage{graphicx}
\usepackage{braket}
\usepackage{color}
\usepackage{multicol}

\begin{document}
\begin{CJK*}{GBK}{song}

\title{Doping-driven orbital-selective Mott transition in multi-band
       Hubbard models with crystal field splitting \thanks{Project
       supported by the National Natural Science Foundation of
       China (Grant No.2011CBA00108) and by the 973 program of China (Grant No.2013CB921700).}}

\author{Yilin Wang$^{1}$, \ Li Huang$^{2}$, \ Liang Du$^{3}$
        \ and \ Xi Dai$^{1}$ \thanks{Corresponding author. E-mail:~daix@iphy.ac.cn} \\
$^1${Beijing National Laboratory for Condensed Matter Physics, }\\
    { and Institute of Physics, Chinese Academy of Sciences, Beijing 100190, China} \\
$^2${Science and Technology on Surface Physics and Chemistry Laboratory,}\\
    { P.O. Box 9-35, Jiangyou 621908, China}\\
$^3${Department of Physics, The University of Texas at Austin, Austin, TX 78712, USA}}

\date{\today}
\maketitle

\begin{abstract}
We have studied the doping-driven orbital-selective Mott transition in multi-band Hubbard models
with equal band width in the presence of crystal field splitting. Crystal field
splitting lifts one of the bands while leaving the others degenerate. We use single-site
dynamical mean-field theory combined with continuous time quantum Monte Carlo impurity solver to
calculate a phase diagram as a function of total electron filling $N$ and crystal field
splitting $\Delta$. We find a large region of orbital-selective Mott phase in the phase diagram
when the doping is large enough. Further analysis indicates that the large region of
orbital-selective Mott phase is driven and stabilized by doping. Such models may account for the
orbital-selective Mott transition in some doped realistic strongly correlated materials.
\end{abstract}



\section{Introduction}
The Mott-Hubbard metal-insulator transition (MIT)\ucite{imada:1039} caused by  electron-electron interaction
is one of the central issues in the physics of strongly
correlated electron systems. A variety of many-body methods such as dynamical mean-field theory
(DMFT)\ucite{antoine:13,kotliar:865} have been developed to study the transitions in models and
realistic materials. Till now, most of the model studies on MIT are focused on the
single band Hubbard model, which has been viewed as the ``Hydrogen atom" for MIT.
However, in most of the realistic materials, the transition should be described by multi-band Hubbard models
and the transition is greatly influenced by the fluctuation of electrons among different orbitals,
which is absent in the simple single band model. As a consequence, the physics of MIT in multi-band systems
is much richer with many exotic phenomena occurring, i.e. the orbital-selective Mott transition (OSMT)\ucite{Anisimov:191}.

The OSMT is first introduced to explain the complicated behavior of conduction electrons in
Ca$_{2-x}$Sr$_x$RuO$_4$\ucite{Anisimov:191}, where at least part of the conduction electrons are itinerant
while the rest seem to be localized. Since then, a lot of model studies have been carried out to uncover the mechanism
of OSMT.
Firstly, a two-band Hubbard model with different band width at half-filling has been investigated
by different groups\ucite{Liebsch:226401,Liebsch:165103,Koga:216402,Koga:045128,Koga:359,Inaba:2393,Medici:205124,
Ferrero:205126,Held:201102(R),Knecht:081103(R)}.
In this case, electrons are equally distributed within the two bands. With the increment of Coulomb interaction $U$,
two different bands undergo MIT separately. The band with narrower band width becomes Mott insulator first due to
its stronger correlation strength while the one with wider band width still remains metallic, indicating an OSMT.
Those studies also confirm that the Hund's coupling, both the Ising-type\ucite{Medici:205124,Knecht:081103(R)}
(only $J_z$ is considered) and the rotational invariant type\ucite{Koga:216402,Koga:045128,Koga:359,Inaba:2393,
 Medici:205124,Ferrero:205126,Held:201102(R)} (including spin-flip $J_s$ and pair-hopping $J_p$ terms),
plays a very important role in OSMT.
Secondly, OSMT can exist in a three-band Hubbard model with equal band width in the presence of crystal field
splitting (CFS) at integer filling 2, 3, or 4. The average filling of four electrons per site was studied in detail
by de'Medici \textit{et al.} in Ref.~\cite{Medici:126401}. In their model, one orbital is lifted by the CFS, while the
other two orbitals keep degenerate. They find a large orbital-selective Mott phase (OSMP) region in the phase diagram and large Hund's coupling
can enlarge the OSMP region as a ``band decoupler".
The case of two electrons per site in a three-band Hubbard model has been studied by Kita \textit{et al.}
in Ref.~\cite{Kita:195130} and they find an OSMT region in the phase diagram at large Hund's coupling.
Thirdly, doping-driven OSMT
has also been studied by some groups\ucite{Inaba:094712,Jakobi:115109,Jakobi:205135,Werner:126405,Werner:115119,Rincon:241105,Rincon:106405}.
In the case of two-band Hubbard model with different band width, OSMT can still be stable away from half-filling
in certain parameters region\ucite{Inaba:094712,Jakobi:115109,Jakobi:205135}.
Werner \textit{et al.}\ucite{Werner:126405} have considered a two-band Hubbard model with equal band width
split by crystal field. They find that, in the presence of CFS, doping would first drive
one of the bands into metallic phase while leaving the other one in a Mott phase, thus inducing the OSMT.
Further doping would drive the other band into metallic phase. In this case, the underlying
physics to drive an OSMT is doping. However, the physics behind such a doping-induced OSMT with equal band width
has not been studied in detail yet, which inspires the current study.

In this paper, we will focus on the detailed physical aspects of doping-driven OSMT.
We consider a multi-band Hubbard model with equal band width, in which the CFS lifts one of the bands to higher energy (upper band)
and leaves the other bands (lower bands) to be degenerate. Through the DMFT calculations, we will show that in such a system
we can find a doping-induced OSMT, which is very surprising because usually doping electrons will turn
the Mott phase into the metallic phase or the opposite.

We start from the situation where the lower bands are half-filled and the upper band is forced to be empty
by adjusting the CFS. Suppose that at the beginning the system is near the critical point of MIT but still
in the metallic side. Then, we start to dope electrons into the upper band but still keep the lower bands
half-filled by tuning the CFS properly.
We find that there will be a critical doping point, after which the lower half-filled bands will be driven into
a Mott phase while the upper band still keeps metallic phase within a considerable region of the CFS.
We also find that further increasing doping concentration will enlarge the OSMP region.
Finally, we reach the critical doping concentration at which both the upper and the lower bands
are half-filled, and the OSMP merges into a single Mott phase. Therefore, we realize a doping-driven
OSMT in multi-band systems in the presence of CFS. The systems described above can be realized
by both two-band and three-band Hubbard models.
In this paper, both models are studied in detail by the DMFT method combined with the high precision
hybridization expansion continuous time quantum Monte Carlo (HYB-CTQMC) impurity solver\ucite{werner:076405,werner:155107,gull:349}.
The phase diagram as a function of total electron filling $N$ and CFS $\Delta$
are calculated.
Based on the results, we summarize this mechanism of OSMT as:
(1) with strong Hund's coupling, doping electrons to a multi-band system with special CFS will help to
form and stabilize the local magnetic moment;
(2) large local magnetic moment will scatter electrons strongly and enhance the effective correlation
strength, which will cause a Mott transition in some certain bands, thus leading to an OSMT.

\section{Model and method}
We consider the following multi-band Hubbard model defined as\ucite{georges:2013},
 \begin{eqnarray}
 \label{eqn:model}
 H &=& -t\sum_{\langle i,j \rangle,\alpha\gamma\sigma} c_{i\alpha\sigma}^{\dagger}c_{j\gamma\sigma}+\sum_{i,\alpha\sigma} (-\mu+\Delta_{\alpha})n_{\alpha\sigma} \nonumber\\
   & &+U\sum_{i,\alpha} n_{\alpha\uparrow} n_{\alpha\downarrow} + \sum_{i,\alpha<\gamma,\sigma \sigma^{\prime}}(U^\prime-\delta_{\sigma\sigma^\prime}J_z) n_{\alpha\sigma}n_{\gamma\sigma^\prime}\nonumber \\
   & & - J_{s} \sum_{\alpha<\gamma, \sigma \neq \sigma^{\prime}} c_{\alpha\sigma}^{\dagger}c_{\alpha\sigma^{\prime}}c_{\gamma\sigma^{\prime}}^{\dagger}c_{\gamma\sigma} + J_{p}\sum_{\alpha\neq\gamma}c_{\alpha\uparrow}^{\dagger}c_{\alpha\downarrow}^{\dagger}c_{\gamma\downarrow}c_{\gamma\uparrow},
 \end{eqnarray}
where $i,j$ is the site index, $\sigma,\sigma^{\prime}=\uparrow,\downarrow$ is the spin index,
$\alpha,\gamma$ is the orbital index and $n_{\alpha\sigma}=c_{\alpha\sigma}^{\dagger}c_{\alpha\sigma}$
is the electron density operator. Here, $t$ is the strength of the nearest-neighbor hopping,
$\mu$ is the chemical potential, and $\Delta_{\alpha}$ is the energy level for orbital $\alpha$.
In the Coulomb interaction terms, $U(U^\prime)$ is the intra-orbital (inter-orbital) Coulomb interaction,
and $J_{z},J_{s},J_{p}$ is the Hund's coupling. The constrained condition $U=U^\prime+2J_{z},J_{z}=J_{s}=J_{p}$ is imposed as usual.
In the present paper, we consider both the Ising-type (only $J_{z}$ term) and the rotational invariant (including $J_{s}$ and $J_{p}$ terms)
Hund's coupling. However, our results show that the physical pictures are almost the same in both cases,
so we will mainly discuss the results of the Ising-type case for simplicity.
The chemical potential $\mu$ is adjusted dynamically in the calculations to reach the expected total electron filling
per site during doping. The CFS is defined as $\Delta=\Delta_{1}-\Delta_{2}(\Delta_1>\Delta_2)$ for two-band case
and $\Delta=\Delta_{1}-\Delta_{2} (\Delta_1>\Delta_{2}=\Delta_{3})$ for three-band case, respectively.

We can solve this lattice model defined in Eqn.~\ref{eqn:model} in the framework of single-site DMFT\ucite{antoine:13,kotliar:865},
which neglects the momentum dependence of the self-energy and maps the lattice model into an effective impurity model being
solved self-consistently. We use a semicircular density of states, $\rho(\omega)=(2/\pi D)\sqrt{1-(\omega/D)^2}$,
which corresponds to the infinite-coordination Bethe lattice, where $D=2t$ is the half band width and all bands keep equal band width.
To solve the effective impurity problem, we use the HYB-CTQMC impurity solver\ucite{werner:076405, werner:155107,gull:349} implemented
in the $i$QIST package\ucite{huangli:2015}, which allows us to access the strong interaction regime down to very low temperature.
The temperature is set to be $T=t/40$ (corresponding to the inverse temperature $\beta t=40$) through all our calculations.
In each DMFT iteration, we have typically performed $2 \times 10^8$ QMC samplings to reach sufficient numerical accuracy.

\section{Results and discussion}

\begin{figure}
\centering
\includegraphics[width=0.50\textwidth]{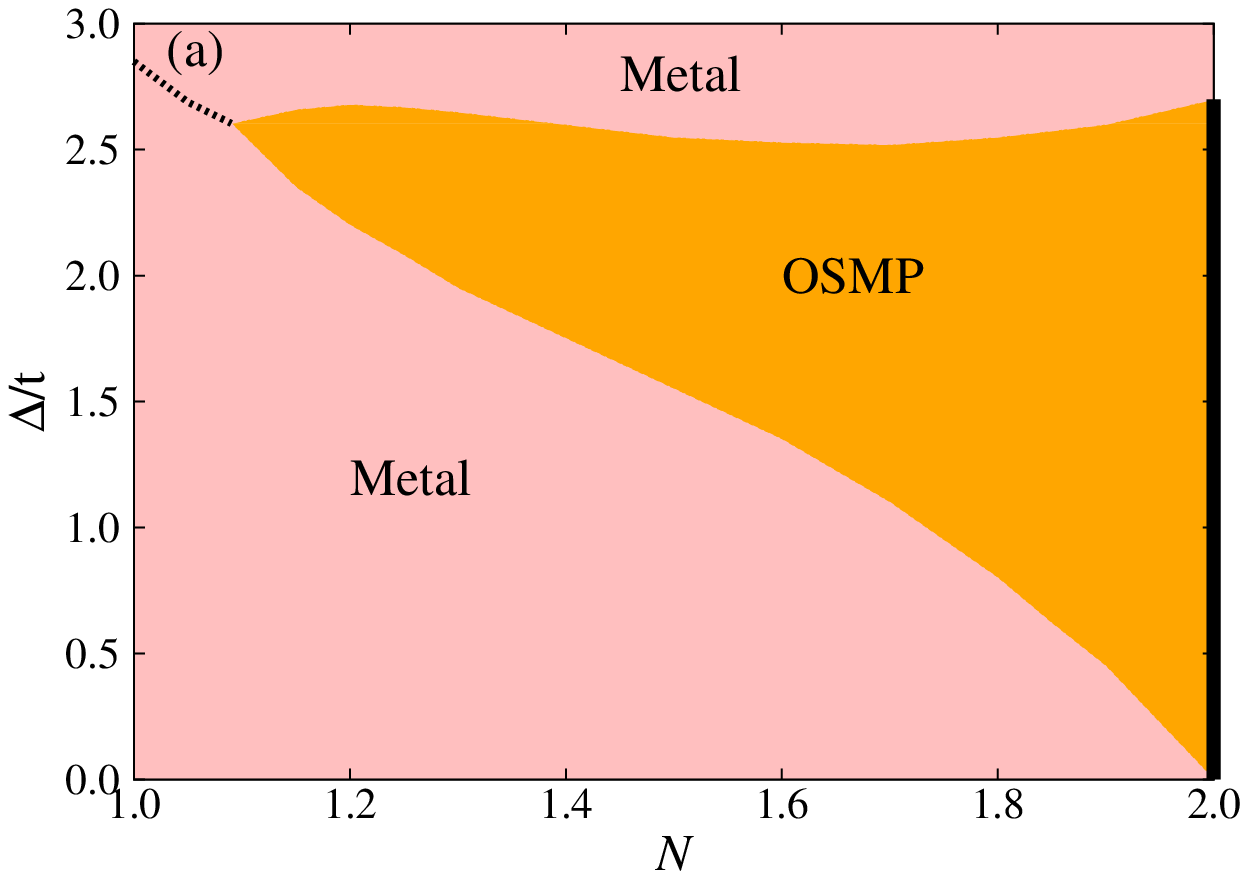}
\includegraphics[width=0.50\textwidth]{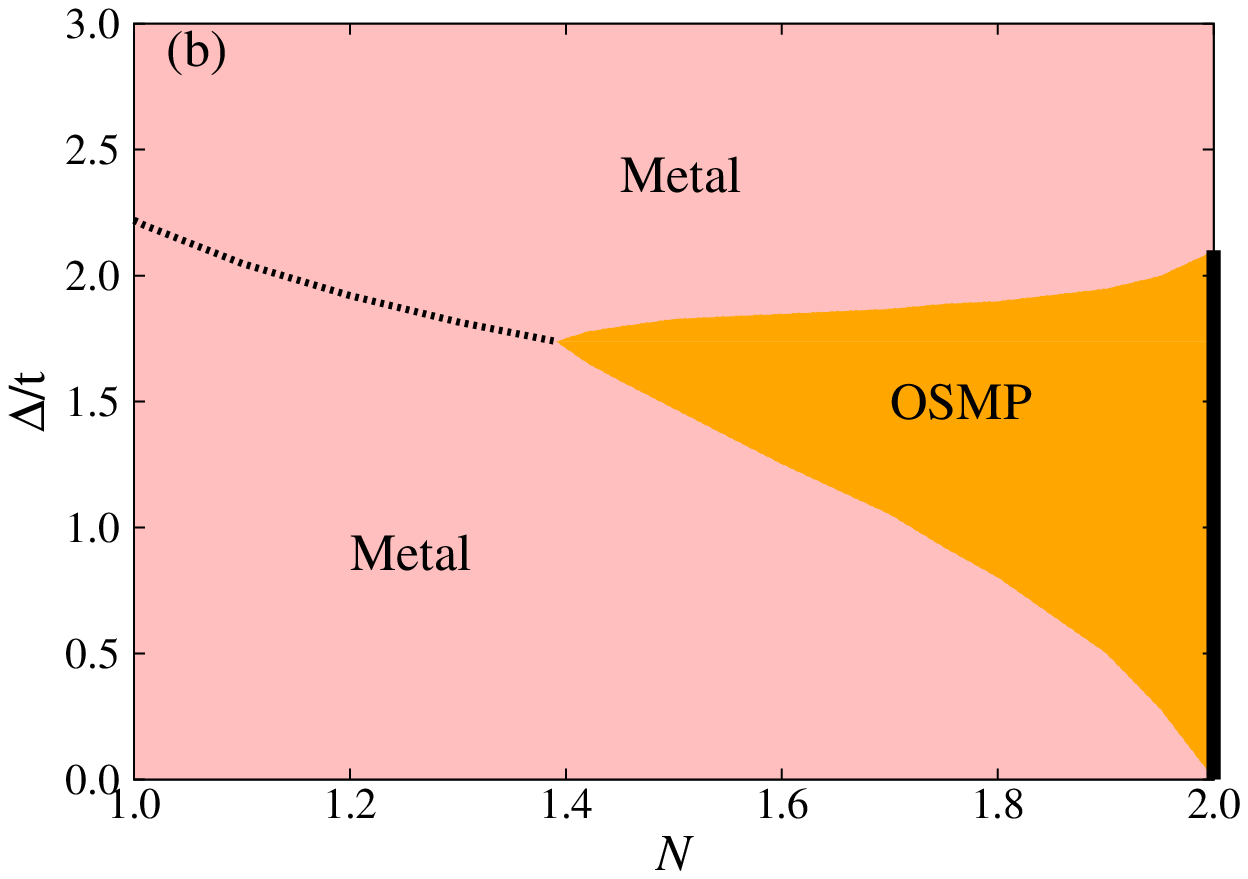}
\caption{(Color online) The phase diagram as a function of the total electron filling $N$ and the crystal field splitting $\Delta$
                        for the two-band Hubbard model with Ising-type Hund's coupling, $U/t=4.90$, (a) $J_{z}=0.25U$ and (b) $J_{z}=0.20U$.
                        The orange area denotes the orbital-selective Mott phase, where the lower orbital
                        is in a Mott phase while the upper orbital is metallic. The pink area denotes that
                        both orbitals are metallic. The black dashed line denotes that the lower orbital is
                        half-filled but still metallic. The black bar at $N=2$ indicates that
                        both orbitals are half-filled and in a Mott phase.}
\label{fig:phase}
\end{figure}

Now, we will first focus on the two-band case.
To demonstrate the OSMT when doping the system, we have calculated the phase diagram as a function of the total electron filling
$N$ and the CFS $\Delta$. The electron filling per orbital per spin, spin-spin correlation function $\chi(\tau)$
and the effective local magnetic moments $M^{z,\text{eff}}$ as a function of the CFS for various $N$ are also calculated to understand
the phase diagram. Based on these results, we will show the importance of doping which drives and stabilizes the OSMT.

The calculated $N-\Delta$ phase diagram is shown in Fig.~\ref{fig:phase}.
The Coulomb interaction is set to be $U/t=4.90$ and two different Hund's coupling values are considered: (a) $J_{z}=0.25U$
and (b) $J_{z}=0.20U$. We choose this value of Coulomb interaction in order to make the system at the critical point of MIT,
and of course, other values of $U$ can also give very similar phase diagram, only with different phase boundary.
We calculate the quasiparticle weight by $Z_{\alpha}^{-1}\approx 1-\frac{\text{Im}\Sigma_{\alpha}(i\omega_0)}{\omega_0}$
to determine the phase for each orbital, where $\omega_0=\frac{\pi}{\beta}$ and $\Sigma_{\alpha}(i\omega_n)$ is the self-energy function.
It is metallic when $Z_{\alpha}$ is finite and it is Mott phase when $Z_{\alpha}$ approaches to zero.
Except for the metallic phase in the pink region, we find a large OSMP region (orange area) and this region
is extended at larger Hund's coupling. The black dashed line indicates the state in which the lower orbital is half-filled
but still metallic and the black bar at $N=2$ indicates that both orbitals are half-filled and in a Mott phase.

\begin{figure}
\centering
\includegraphics[width=0.50\textwidth]{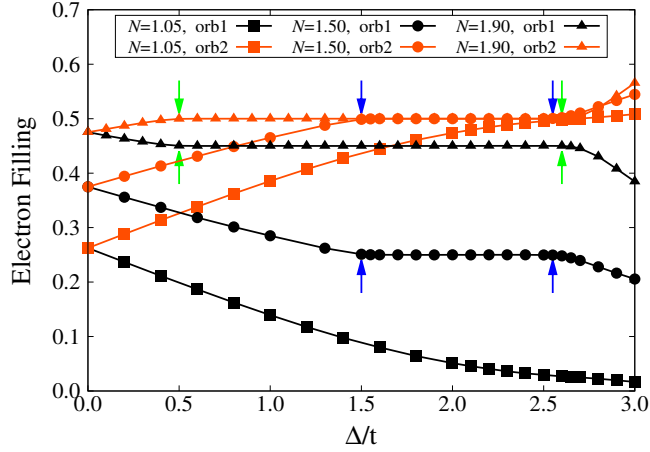}
\caption{(Color online) The electron filling per orbital per spin as a function of the
                        crystal field splitting $\Delta$ for three selected total electron filling
                        $N=1.05, 1.50, 1.90$, respectively. $U/t=4.90, J_{z}=0.25U$. Note that the upper orbital is
                        marked as ``orb1" and the lower orbital is marked as ``orb2".
                        }
\label{fig:occu}
\end{figure}

To understand the phase digram, we plot the redistribution of electrons when increasing the CFS
at a fixed total electron filling $N$, the results for $U/t=4.90, J_{z}=0.25U$ are shown in Fig.~\ref{fig:occu}.
We take $N=1.50$ for example. Initially, in the absence of the CFS, the electrons distribute equally
within the two orbitals. If we continuously increase $\Delta$, the electrons will gradually populate
into the lower orbital until it is half-filled (indicated by the left blue arrow in Fig.~\ref{fig:occu}).
After that, further increasing the CFS will not populate more electrons into the lower orbital
because it will pay a higher energy cost due to the strong Coulomb interaction among the electrons in the lower orbital.
As a result, the system may come into a state, in which the lower orbital always keeps half-filling (the plateau between
the blue arrows in Fig.~\ref{fig:occu}) in a considerable region of the CFS.
We also find that this plateau will reach earlier at large doping (see $N=1.90$ for example, plateau indicated by the green arrows).
On the contrary, for very small doping (for example $N=1.05$), we cannot see obvious plateau.
Therefore, the plateau becomes wider when doping is larger.
However, if the CFS is large enough (comparable with $U$), the electrons will again populate into the lower
orbital to avoid the cost of the large CFS energy.

\begin{figure}
\centering
\includegraphics[width=0.50\textwidth]{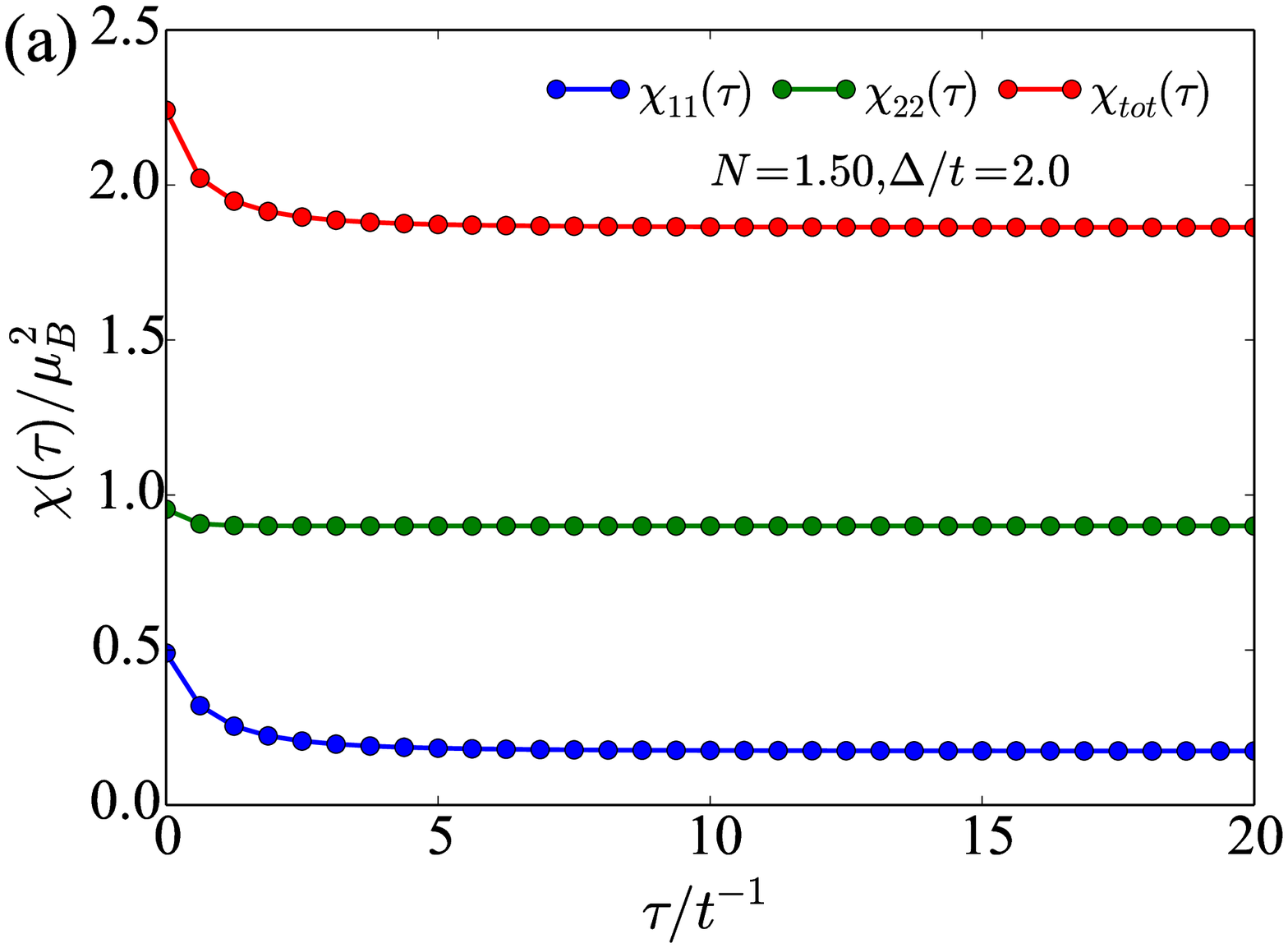}
\includegraphics[width=0.50\textwidth]{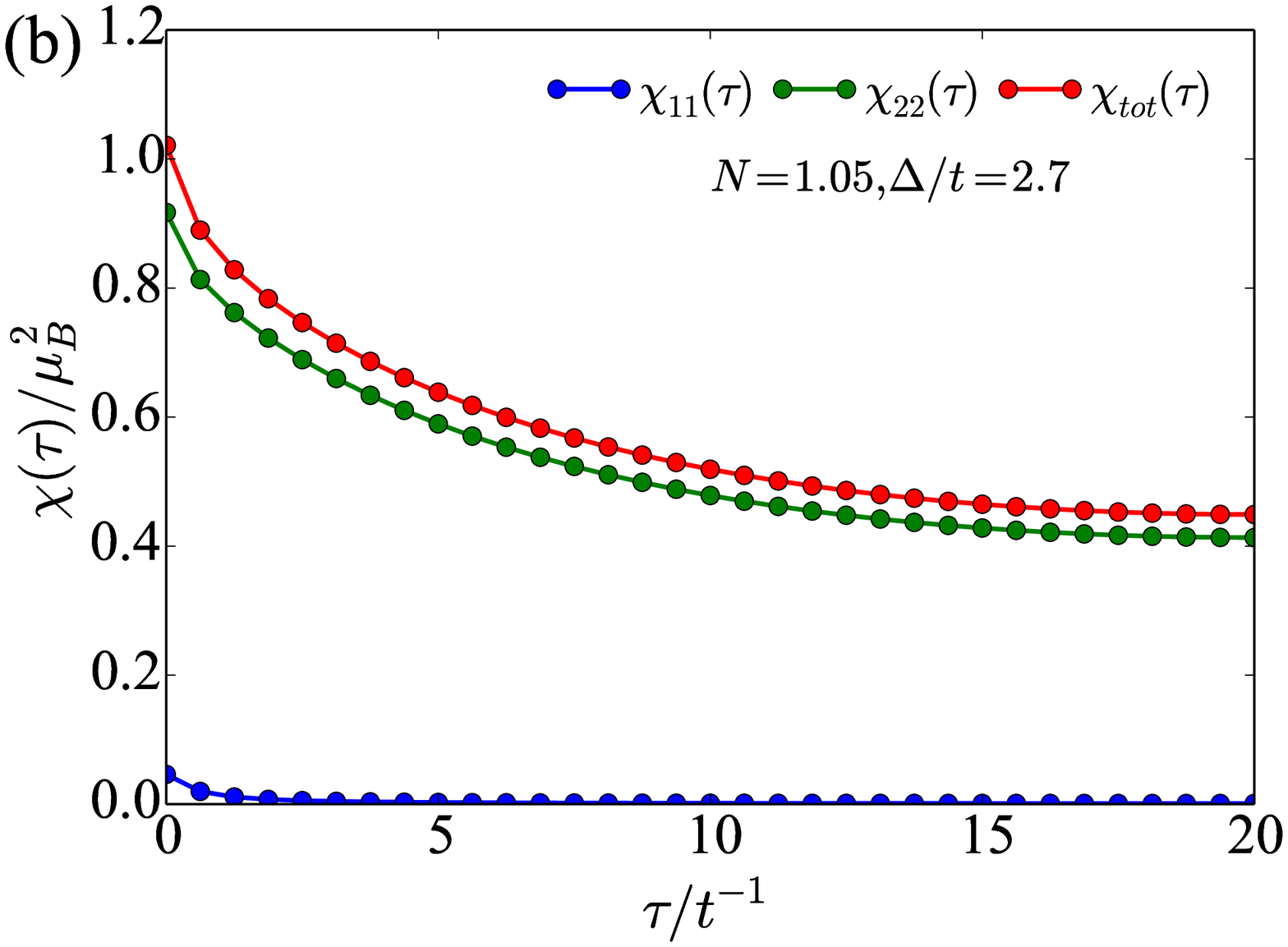}
\caption{(Color online) The spin-spin correlation function $\chi(\tau)$ as a function of imaginary time $\tau$ from
         0 to $\beta/2$ at $U/t=4.90, J_{z}=0.25U$. (a) $N=1.50, \Delta/t=2.0$, in OSMP phase,
         (b) $N=1.05, \Delta/t=2.7$, in metallic phase. In both cases, the lower orbital is half-filled.
          $\chi_{11}(\tau), \chi_{22}(\tau), \chi_{\text{tot}}(\tau)$ are for upper orbital, lower orbital and
          total contributions, respectively.}
\label{fig:chitau}
\end{figure}

\begin{figure*}
\centering
\includegraphics[width=1.00\textwidth]{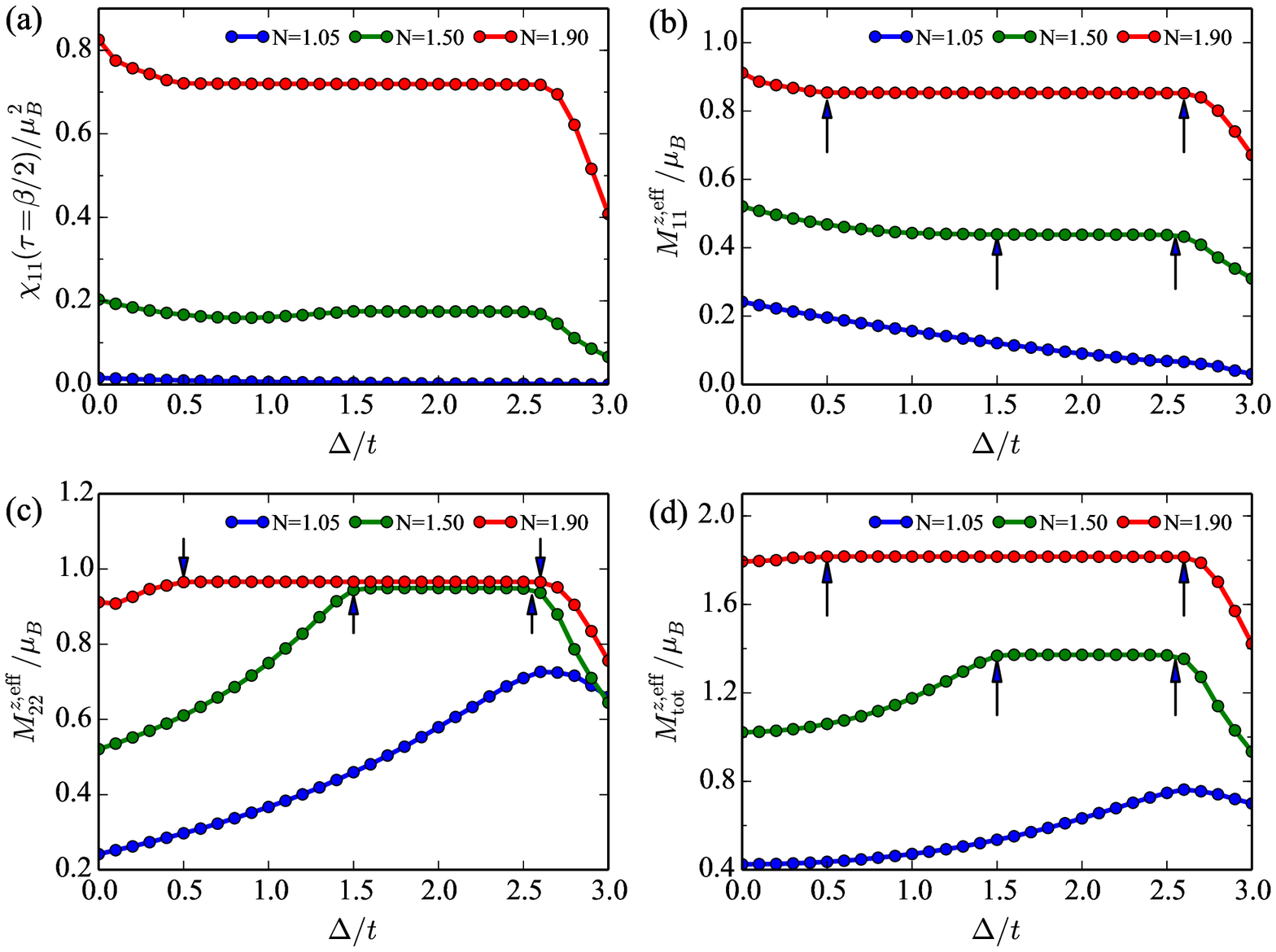}
\caption{(Color online) Spin-spin correlation function and effective local magnetic moments as a function of
         crystal field splitting $\Delta$ for $N=1.05, 1.50, 1.90$ at $U/t=4.90, J_{z}=0.25U$.
         (a) The spin-spin correlation function at $\tau=\beta/2$ for upper orbital $\chi_{11}(\tau=\beta/2)$,
         (b) The effective local magnetic moments for upper orbital $M^{z,\text{eff}}_{11}$,
         (c) The effective local magnetic moments for lower orbital $M^{z,\text{eff}}_{22}$,
         (d) The total effective local magnetic moments $M^{z,\text{eff}}_{\text{tot}}$.
         The arrows indicate plateaus.}
\label{fig:meff}
\end{figure*}

Based on the knowledge about the redistribution of electrons discussed above, we can understand
the whole phase diagram. The metallic phase in pink region can be understood easily because both
of the orbitals are away from the condition of half-filling.
The most interesting part of the phase diagram is the metallic phase indicated by the black dashed line
and the OSMP in the orange region.
The black dashed line represents the state in which the lower orbital is half-filled but metallic.
On the contrary, in the orange area, the lower orbital is also half-filled but it is in a Mott phase,
and as a whole the system is in an OSMP. These facts remind us that the lower half-filled orbital should have undergone
a MIT when the doping becomes larger and the MIT has strong connection with the plateau of electron filling.
Actually, as mentioned above, when the filling of the lower orbital continues to
increase to be half-filled by increasing the CFS, the rest of the doped electrons will fill the upper orbital,
and these electrons will interact with the electrons in the lower orbital through the strong Hund's coupling
to help to form total local magnetic moments and stabilize them. The more doped electrons, the larger magnetic moments and it means that the system favors
a high-spin state (HS)\ucite{Werner:126405} when doping becomes large.

To illustrate this, we calculate the spin-spin correlation function and the effective local magnetic moments.
The spin-spin correlation function\ucite{werner:2008,Haule:025021,Toschi:064411} can be used to describe the dynamical
screening of the local magnetic moments, which is defined as,
\begin{equation}
  \chi_{\text{tot}}(\tau) = \sum_{\alpha\gamma}\chi_{\alpha\gamma}(\tau) = \sum_{\alpha\gamma}(g\mu_{B})^2\Braket{S_{z}^{\alpha}(\tau)S_{z}^{\gamma}(0)},
\end{equation}
where, $0\leq\tau\leq\beta$ is the imaginary time, $S_{z}^{\alpha}=\frac{1}{2}(n_{\alpha\uparrow}-n_{\alpha\downarrow})$ is the total $z$-component of
spin angular momentum for orbital $\alpha$, $g=2$ is the Land\'e $g$-factor for spin, $\mu_{B}$ is the Bohr magnetic moment.
$\chi(\tau=0)$ measures instantaneous magnetic moments formed mainly due to the Hund's coupling and the screened magnetic moments
can be measured at longer time $\tau$ ($\sim\beta/2$), so if $\chi(\tau)$ approaches to a nonzero constant at longer time,
then it indicates that some screened (frozen) moments\ucite{werner:2008,Toschi:064411} survive after the dynamical screening by the mobile electrons.
In figure~\ref{fig:chitau}(a,b), we plot $\chi(\tau)$ for upper orbital $\chi_{11}(\tau)$, lower orbital $\chi_{22}(\tau)$ and total contribution $\chi_{\text{tot}}(\tau)$
for two cases $N=1.50, \Delta/t=2.0$ and $N=1.05, \Delta/t=2.7$ at $U/t=4.90, J_{z}=0.25U$, in which both the lower orbitals are half-filled.
For $N=1.50, \Delta/t=2.0$, the system is in OSMP, and $\chi(\tau)$ of both orbitals quickly approaches to
a nonzero constant, which indicates that the screening is not effective and not only well-defined local moments form in the half-filled insulating band
but also frozen local moments form in the partially filled metallic band.
For $N=1.05, \Delta/t=2.7$, both orbitals are metallic, and $\chi(\tau)$ quickly decays and approaches to a small constant
at very long time, which means that the screening process is much more effective and only small frozen moments survive.
So, we can conclude that increasing doping indeed helps to form local magnetic moments.
To further measure the total and orbital-resolved local magnetic moments quantitatively, we define the effective $z$-component
of local magnetic moments $M^{z,\text{eff}}$ by averaging $\chi(\tau)$ over $\tau$ from 0 to $\beta$, which are
\begin{equation}
M^{z,\text{eff}}_{\alpha\gamma} = g\mu_{B}\sqrt{\frac{1}{\beta}\int_{0}^{\beta}\Braket{S_{z}^{\alpha}(\tau)S_{z}^{\gamma}(0)}\mathrm{d}\tau}.
\end{equation}
\begin{equation}
M^{z,\text{eff}}_{\text{tot}} = g\mu_{B}\sqrt{\frac{1}{\beta}\int_{0}^{\beta}\sum_{\alpha\gamma}\Braket{S_{z}^{\alpha}(\tau)S_{z}^{\gamma}(0)}\mathrm{d}\tau},
\end{equation}
Note that the Curie-Weiss law $(M^{z,\text{eff}}_{\text{tot}})^{2}=(g\mu_{B})^2S(S+1)/3$ will be satisfied if $\chi(\tau=\beta/2)$
is very close to $\chi(\tau=0)$ and the system is rotationally invariant, where $S$ is the quantum number of total spin.
In our case, the system is not rotationally invariant, so we just discuss the $z$-component of them as an approximation to measure the size of local magnetic moments.
In figure~\ref{fig:meff}, we plot the spin-spin correlation function at $\tau=\beta/2$ for upper orbital $\chi_{11}(\tau=\beta/2)$ (Fig.~\ref{fig:meff}a),
the effective local magnetic moments for upper orbital $M^{z,\text{eff}}_{11}$ (Fig.~\ref{fig:meff}b),
for lower orbital $M^{z,\text{eff}}_{22}$ (Fig.~\ref{fig:meff}c) and for total contribution $M^{z,\text{eff}}_{\text{tot}}$ (Fig.~\ref{fig:meff}d)
as a function of CFS for $N=1.05, 1.50, 1.90$ at $U/t=4.90,J_{z}=0.25U$, respectively.
For large doping, the upper metallic orbital has considerable frozen local magnetic moments which
will contribute to the total local magnetic moments, see Fig.~\ref{fig:meff}(a,b).
In the OSMP phase for $N=1.50, 1.90$, there is a plateau indicating large effective local magnetic moments formed by the lower Mott insulating orbital, see Fig.~\ref{fig:meff}(c).
There are large well-defined total effective local magnetic moments $M^{z,\text{eff}}_{\text{tot}}$ in OSMP phase, which
are labeled by the plateaus at large doping $N=1.50, 1.90$ in Fig.~\ref{fig:meff}(d).
Note that, by definition, $M^{z,\text{eff}}_{\text{tot}} \neq M^{z,\text{eff}}_{\text{11}} + M^{z,\text{eff}}_{\text{22}}$,
instead, $M^{z,\text{eff}}_{\text{tot}} = \sqrt{(M^{z,\text{eff}}_{11})^2+(M^{z,\text{eff}}_{12})^2+(M^{z,\text{eff}}_{21})^2+(M^{z,\text{eff}}_{22})^2}$.

In the presence of large local magnetic moments, the charge fluctuation will be suppressed
because the electrons will be scattered by the local magnetic moments. As a result, the effective
Coulomb correlation strength $U_{\text{eff}}$ of the lower half-filled orbital will be enhanced when increasing doping,
after a critical value $U_{\text{eff}}^{c}$, the lower orbital will undergo a MIT and an OSMT occurs.
In summary, the more doping, the larger local magnetic moments and the wider OSMP region,
and the OSMT occurs as long as the electron filling per orbital comes into the plateau.
In the OSMP region, the itinerant electrons of the upper orbital will show
non-Fermi liquid (NFL) behavior\ucite{werner:2008}. This is due to the scattering by the large total local magnetic moments,
including the contribution of local moments in the lower insulating band and the frozen moments in the upper metallic band.
At $N=2.0$, the upper orbital also undergoes a MIT, as a whole, the system is in the Mott phase,
which is illustrated as the black bar in Fig.~\ref{fig:phase}.
Here, we should emphasis that the Hund's coupling plays a crucial role.
We can see that the OSMP region for $J=0.20U$ is smaller than that for $J=0.25U$.
Thus, large Hund's coupling is a prerequisite for OSMT, which is consistent with the conclusion of previous studies\ucite{Medici:205124,
Knecht:081103(R),Koga:216402,Koga:045128,Koga:359,Inaba:2393,Medici:205124,Ferrero:205126,Held:201102(R)}.

\begin{figure}
\centering
\includegraphics[width=0.50\textwidth]{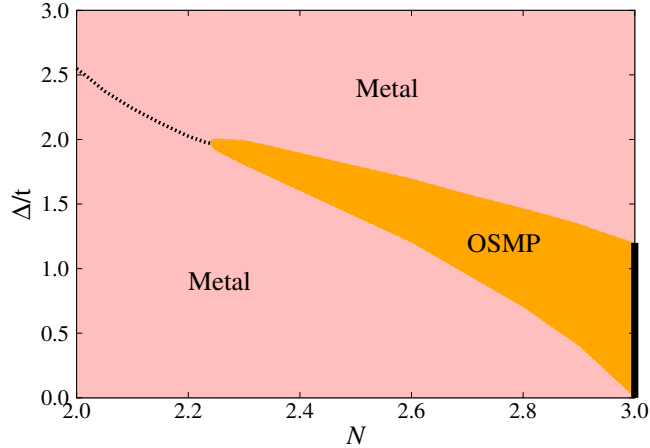}
\caption{(Color online) The phase diagram as a function of the total electron filling $N$ and the crystal
         field splitting $\Delta$ for the three-band Hubbard model with
         Ising-type Hund's rule coupling at $U/t=2.80, J_{z}=0.25U$. The orange area denotes the orbital-selective
         Mott phase where the lower two orbitals are in a Mott phase while the upper orbital is
         metallic. The pink area denotes that all orbitals are metallic. The black dashed line indicates
         that the lower two orbitals are half-filled but still metallic. The black bar at $N=3$
         indicates that all the three orbitals are half-filled and in a Mott phase. }
\label{fig:phase_3bnd}
\end{figure}

Next, we discuss the case of three-band Hubbard model with Ising-type Hund's coupling.
In this case, two orbitals keep degenerate and another orbital is lifted by the CFS.
The $N-\Delta$ phase diagram at $U/t=2.80,J=0.25U$ is shown in Fig.~\ref{fig:phase_3bnd}.
The main character of this phase diagram is very similar with that of the two-band model and the
mechanism of OSMT doesn't change. The OSMT occurs when the lower two orbitals are half-filled
and the doping is large enough. It seems that the OSMP region is smaller than that of
the two-band case, actually, this is due to the choice of the Coulomb interaction $U$ and Hund's coupling $J_{z}$.

Finally, we would emphasize that the difference between this mechanism of OSMT and previous
studies\ucite{Inaba:094712,Jakobi:115109,Jakobi:205135} of the OSMT in doped multi-band Hubbard models.
In their studies, the bands have different band width, so the effective Coulomb interaction strength
is different for each band, which may lead to an OSMT.
However, in our study, we don't need the condition of different band width, and the
effective Coulomb interaction strength of the lower bands will be enhanced by the local magnetic moments
which are formed and stabilized by the doping electrons of the upper band, thus leading to an OSMT.

\section{Conclusion}
\label{sec:con}
In summary, we have found a doping-driven OSMT in multi-band Hubbard model with equal band width
in the presence of the CFS. When the Hund's coupling is large enough, the doped electrons in the upper
orbital will help to form large local magnetic moments which will scatter electrons and suppress the charge fluctuation,
thus the effective correlation strength of the lower orbitals will increase large enough to undergo a MIT. In such a way,
an OSMT occurs. The more doping, the more stable OSMP. This mechanism of OSMT can be realized in both two-band
Hubbard model and three-band Hubbard model with Ising-type and rotational invariant Hund's coupling.
We hope that this mechanism of OSMT can also be found in some realistic strongly correlated metallic materials.

\addcontentsline{toc}{chapter}{Acknowledgment}
\section*{Acknowledgment}
We thank useful discussion with Philipp Werner.
The DMFT+CTQMC calculations have been performed on the cluster of DAWNING at Institute of Physics, Chinese Academy of Sciences.

\end{CJK*}
\end{document}